\documentclass[printer]{aa}
\usepackage{graphicx}
\usepackage[]{natbib}
\begin{document}

\title{The radial metallicity gradient and the history of elemental enrichment in M81 through emission-line probes}
\subtitle{}

\author{Letizia Stanghellini
\and Laura Magrini
\and Viviana Casasola
\and Eva Villaver
}
 
\institute{
National Optical Astronomy Observatories, Tucson, AZ 85719;
\email{lstanghellini@noao.edu}
\and
INAF - Osservatorio Astrofisico di Arcetri, Largo E. Fermi, 5, I-50125 Firenze, Italy;
\email{laura@arcetri.astro.it}
\and
INAF - Istituto di Radioastronomia \& Italian ALMA Regional Centre, Bologna, Italy;
\email{casasola@ira.inaf.it}
\and
Universidad Aut\'onoma de Madrid, Departamento de F\'isica Te\'orica C-XI, 28049 Madrid, Spain
\email{eva.villaver@uam.es}
}

\date{}

\abstract
 {}
   { We present a new set of weak-line abundances of HII regions in  M81, based on Gemini Multi-Object Spectrograph (GMOS) observations.  The aim is to derive plasma and abundance analysis for a sizable set of emission-line targets to study the galactic chemical contents in the framework of galactic metallicity gradients.}
   {We used the weak-line abundance approach by deriving electron density and temperatures for several HII regions in M81. Gradient analysis is based on oxygen abundances.}
   {Together with a set of HII region abundances determined similarly by us with Multi-Mirror Telescope (MMT) spectra, the new data yield to a radial oxygen gradient of -0.088$\pm$0.013 dex kpc$^{-1}$, which is steeper than the metallicity gradient obtained for planetary nebulae (-0.044$\pm$0.007 dex kpc$^{-1}$). This result could be interpreted as gradient evolution with time:  Models of galactic evolution with inside-out disk formation associated to  pre-enriched gas infall would produce such difference of gradients, although stellar migration effects would also induce a difference in the metallicity gradients between the old and young populations. }
 {By comparing the M81 metallicity gradients with those of other spiral galaxies, all consistently derived from weak-line analysis, we can infer that similar gradient difference is common among spirals.  The metallicity gradient slopes for HII regions and PNe seem to be steeper in M81 than in other galactic disks, which is likely due to the fact that M81 belongs to a galaxy group. We also found that M81 has experienced an average oxygen enrichment of 0.14$\pm$0.08 dex in the spatial domain defined by the observations. Our data are compatible with a break in the radial oxygen gradient slope around R$_{25}$ as inferred by other authors both in M81 and in other galaxies.}


\authorrunning{Stanghellini et al.}
\titlerunning{HII regions in M81}
\maketitle

\section{Introduction}

The metallicity of a galaxy carries the signature of its history, including various phenomena such as gas accretion during the first epochs (infall), star formation, and subsequent gas outflow/inflow.  Among the various observational constraints that can shed light on the galaxy past, an important one is the radial metallicity gradient-- and its evolution with time-- which is sensitive to the assembly history at different radii, and thus tells a story about galaxy formation and evolution processes. 

From an observational point of view, during the last decades the study of the gradient evolution has been mainly investigated through the determination of the metallicity of resolved populations with different ages in the Milky Way and in nearby galaxies. Measurements of metallicity of different targets in the Local Universe (for instance young OB stars and HII regions, Cepheids, open clusters, red giant stars) have shown that disk galaxies usually  exhibit  negative radial metallicity gradients, with higher metallicity in their inner regions and lower metallicity at larger galactocentric radii (e.g., Vila-Costas \& Edmunds 1992; Zaristisky et al. 1994; Rupke et al. 2010).  It was also found that, in several cases, the radial gradient becomes flatter at large radii (e.g., Werk et al. 2011; Bresolin et al. 2012; CALIFA survey results described in Sanchez et al. 2013). Very recently, more attention has also been devoted to measurements of the time evolution of metallicity gradients and its consequent implications for galaxy formation and evolution.
 
In the Galaxy, two stellar tracers have been essentially used to investigate the time evolution of the radial gradient: open clusters  and planetary nebulae (PNe). Open clusters represents a reliable approach to the study of the time evolution of the metallicity gradient, since it is possible to firmly determine their age, Galactocentric distances, and  abundances of a large number of elements (see, e.g., Janes 1979; Freeman \& Bland-Hawthorn 2002; Friel et al. 2002; Magrini et al. 2009a; Bland-Hawthorn et al. 2010; Kobayashi \& Nakasato 2011, Yong et al. 2012). From the recent high spectral resolution studies of open clusters in the inner Galaxy (R$_{\rm G}<$ 13~kpc) the older open clusters show a steeper abundance gradient  than the younger clusters, implying thus a slight gradient flattening with time in the inner Galaxy. 

Planetary nebulae in principle should allow to detect the time evolution of the gradient by comparing the present-time gradient, as outlined by HII regions, with that of PNe of different ages.  In our Galaxy however the distances to PNe are affected to large uncertainties that restrict their use as tracers of the past evolution of metallicity gradients. The best distance scale available to date is the one calibrated on the Magellanic Cloud PNe observed with the {\it Hubble Space Telescope (HST)} (Stanghellini et al. 2008). This distance scale is very similar to Cahn et al. (1992)'s scale, the most commonly used to date. Stanghellini \& Haywood (2010) used Stanghellini et al. (2008)'s scale and oxygen abundances from weak-line analysis to determine Galactic metallicity gradients with PNe of different progenitor ages. The sample of PNe analyzed by Stanghellini \& Haywood (2010) had been parsed into age bins based on their location with respect to the Galactic plane, their peculiar radial velocity, and their nitrogen and helium contents. In fact, the mass range of PN progenitors can be constrained by comparing the observed N and He enrichment to the yields from stellar evolution; this allows to mark the time of PN progenitor formation, and consequently allows the determination of chemical enrichment when comparing $\alpha$-element abundances in young and old populations. The nebular parameters listed above have typically low uncertainties, and are only marginally dependent on assumptions.  Stanghellini \& Haywood (2010) found a very mild steepening of the gradient with time, but the evolution is not significant given the distance scale uncertainties. On the other hand, Maciel and Costa (2013) determined the radial metallicity gradients of several PN populations  with Cahn et al. (1992)'s distance scale, and by using the PN central star properties to date the PN populations; they found radial gradients almost invariant with time, with differences consistent with the age-metallicity dispersion. The different results by the two PN teams could be ascribed to uncertainties associated with dating PN based on central stars -- given that stellar progenitors could have had a non-conventional evolution such as common envelope binaries -- rather than to the distance scale used.

In external galaxies, where PNe can be assumed, to first order, at the same distance of the host galaxy,  the comparison of PNe and HII region abundances, investigated with similar observational and analysis techniques, has given reliable results (e.g., Magrini et al. 2007, 2010; Stanghellini et al. 2010; Stasinska et al. 2013). The large database of PNe and HII regions in M33 indicate that the galaxy underwent a global enrichment similar at all radii, and that the slope of the gradient was essentially unvaried (Magrini et al. 2010). For NGC~300, Stasinska et al. (2013) found that the formal abundance gradients of PNe are shallower than for HII regions. However, their large observed abundance dispersion and the small statistics on which their results are based make any conclusion on a possible steepening of the gradients with time only tentative. 

An alternative approach to the study of the time evolution of metallicity gradients is to derive them at different cosmic epochs, and to compare with local $z$=0 galaxies, taking care of considering galaxies with equivalent dark matter halos.  Some pioneering studies have found unexpected results that suggest that some massive galaxies may show positive gradients with lower metallicity in the central regions (Cresci et al. 2010; Queyrel et al. 2012) at very high redshift ($z$$\sim$3). These galaxies, however, are likely progenitors of present time massive elliptical galaxies and should be not compared with present disk galaxies.  Other recent studies were instead based on the analysis of high-$z$ lensed galaxies:  Jones at al. (2010) measured the metallicity gradient of a gravitationally lensed galaxy at $z$=2  finding it significantly steeper than in local disk galaxies. A similar result was obtained in another lensed galaxy at $z$=1.5 (Yuan et al. 2011).   Other four lensed galaxies were studied very recently by Jones et al. (2013) who compared them with  galaxies at lower redshift selected to occupy equivalent dark matter halos. They found that, on average, gradients  flatten by a factor of $\sim$2.6 between $z$=2.2 and $z$=0, in agreement with size evolution measured for more massive galaxies by van Dokkum et al. (2010).

The emerging scenario at this time is that several Local Universe observations have shown the presence of metallicity gradients in old populations, such as PNe and open clusters. Since radial migration is able only to flatten the gradient redistributing the stars of different ages (see, e.g., Fig.~2 of Roskar et al 2008; Michev et al. 2013), the observations of non null gradients in old population imply that the effect of radial migration of stellar population is not so strong to cancel them. To date, we do not have a firm conclusion on the evolution of the gradients in the Local Universe since different authors found discordant results. However we can conclude that all results are in the framework of a limited evolution of the slope with time. From the high-$z$ Universe observations, the new set of observations of lensed galaxies (consistent to be  present disk-galaxy progenitors) show steeper gradients than that observed presently in disk galaxies. If we compare the high-$z$ results with Fig.~2 in Roskar et al. (2008), we can see that they are not inconsistent with what observed with old stellar population in Local Universe galaxies. 

From a theoretical point of view, different types of classical chemical evolution models --where "classical" means that they do not consider the cosmological context, and they do not consider dynamical effects--predict different temporal behaviors of the metallicity gradients due to the different  rates of the chemical enrichment in inner and outer regions of the galactic disk related to the star formation and infall processes. The models can be broadly divided in  those where the metallicity gradients steepen with time (Chiappini et al. 1997; Chiappini et al. 2001)  and those where they flatten with time (Moll{\'a} et al. 1997; Portinari \& Chiosi 1999; Bossier \& Prantzos 1999; Hou et al. 2000; Magrini et al. 2007, 2009b). 
However, classical models have been recently overcame by the development of models of formation and evolution of galaxies created in a cosmological context (see, e.g.,  Rahimi et al. 2011, Pilkington et al. 2012, Gibson et al. 2013) and by models that join chemical evolution with dynamical aspects (see. e.g., Michev et al. 2013). Only recently these new set of models reached a spatial resolution able to investigate the shape of the metallicity gradients and its evolution with time. 

The conclusions given by Gibson et al. (2013) are instructive of the present time situation. With their modeled galaxies, realized with different assumptions (e.g., with different feedback implementations) they can obtain both gradients that only mildly steepen with time, and metallicity gradients steeper at high redshift, that subsequently flatten with time. 
Both results have an observational counterparts, and thus they conclude that more constraints from the local and high-redshift Universe are necessary to provide more definitive conclusions on the time evolution of the gradients. To a similar conclusions arrives Moll{\'a} et al. (2014), where spectro-photometric models show a moderate flattening of the radial gradients with decreased redshift, flattening that tend to be less noticeable when only the inner parts of the galaxies are accounted for. 

Aiming at adding more constraints to the evolution of radial metallicity gradients, we embarked in observing a significant sample of HII regions in M81, a nearby (3.63$\pm$0.34 Mpc, Freedman et al. 2001) spiral galaxy whose membership to a tidal group is evident from its extended tidal streams, clearly observed in the HI emission line (Gottesman \& Weliachew 1975; Yun et al. 1994). Several tools are available to understand the chemical evolution of the M81 disk: star clusters (Ma et al. 2005, Nantais et al. 2011), young supergiant stars (Davidge 2006), X-ray sources (Sell et al. 2011), and color-magnitude diagram fitting to the HST-resolved stellar population (Kudritzki et al., 2012).  Nonetheless, emission-line targets such as  HII regions and PNe have advantages over their stellar counterparts since PNe and HII region spectra are analyzed in similar ways, making the comparison of the two sets of probes more direct than when comparing sources of different nature.  HII regions and PNe represent  different star formation epochs in the galaxy evolutionary history,  HII regions probing the stellar population currently formed, and PNe being the gaseous remnants of stars formed 1-10 Gyr ago; studied together the two populations provide the temporal dimension of galactic evolution. 
The study of chemical evolution through emission-line probes in M81 has been attempted before. Garnett \& Shields (1987) have used strong-line abundances to constrain the radial metallicity gradient of the M81 disk with oxygen abundances of HII regions, finding a negative gradient of about -0.08 dex kpc$^{-1}$ at intermediate (4-12 kpc) galactocentric radii. The strong-line method provides estimate of elemental abundances, while the weak-line method utilizes auroral line strengths to determine electron temperatures directly, and it is much more accurate in determining ionic abundances. M81 has a well-defined, shallow PN metallicity gradient from weak-line analysis (Stanghellini et al. 2010);  several HII regions have been studied within the same paper as well, but the characterization of the radial metallicity gradient has not been possible given the limited number of probes observed.

While direct comparison of emission line abundances from probes of different progenitors has been attempted successfully in M33 and NGC 300, to date there are no other nearby spirals where adequate spectroscopy is available for both PNe and  HII regions to determine weak-line abundances. This paper represents a continuation of these abundance studies started by us with M33  (Magrini et al. 2009b, 2010), and M81 (Stanghellini et al. 2010). Here, we present the weak-line abundances from a new dataset of HII region spectroscopy in the inner 12 kpc of M81. We observed two M81 fields with the Gemini Multi-Object Spectrograph (GMOS) on Gemini North with the aim of obtaining weak-line abundances for HII regions that would define the radial metallicity gradient in the inner parts of M81, with the main goal to study their metallicity gradient and chemical enrichment. 

We present the data acquisition and analysis in $\S$2; the radial metallicity gradients and metal enrichment of M81 based on our data are in 
$\S$3; the discussion,  in $\S$4, includes a comparison of M81 weak-line abundance gradients with those of other galaxies. The conclusions are given in $\S$5. 

\section{Data acquisition and analysis}

Progress in understanding galaxy formation and evolution is made by constraining evolutionary models with the best available data sets. The chemical evolution of galactic disks in particular is well described by their radial metallicity gradients: abundances of resolved targets, together with their spatial distribution, set solid constraints on the star formation history and rates of the studied galaxies. Abundance studies in nearby galaxies, where the stellar populations can be resolved, offer the opportunity to constrain models of disk evolution which can be then extrapolated to  higher redshifts, where the effects of variations in star formation history and stellar ages are more difficult to quantify. 

Our data set consists of narrow-band and continuum images, and Multi-Object Spectrograph (MOS) spectroscopy, of two M81 fields. Pre-imaging was acquired with GMOS-N in queue mode, while MOS observations were acquired on January 24, 2012 in classical mode from Mauna Kea. A log of all observations is given in \ref{table:1}, where we report the field, date, set up, and, for the queue observations, and the observing conditions (Image Quality, Cloud Cover, and Background), as defined by Gemini.

\subsection{Pre-imaging}

With program GN-2011B-Q-32 we studied two fields of M81, both covering an area of 5.5 arcmin$^2$, located at the periphery (field 1: RA 09:55:16.96; Dec 69:11:19.42) and near the center (field 2: RA 09:55:04.41; Dec 69:05:50.31) of the galaxy. Images were acquired through the H$\alpha$ and corresponding continuum filter, with details in Table 1. 

In Figure 1 we have indicated the location of our fields. Therein, the squares indicate the position of the identified HII regions that we have then observed spectroscopically, and whose spectra have sufficient S/N ratio ($\sim$3) for their lines to be listed in the flux tables, while crosses correspond to targeted regions whose spectra were not well characterized. The regions had been identified in the subtracted images, where the  continuum  images have been scaled then subtracted from the H$\alpha$ images. We have used the GMOS mask-making software ({\it gmmps}, available on Gemini webpages) to produce the two masks for spectroscopy. We endeavored to select only HII regions, avoiding targets that were likely to be planetary nebulae (the slightly fainter, compact H$\alpha$ sources). 

We could perform MOS spectroscopy of 27 HII regions in field 1 and 29 regions in field 2.  

\subsection{MOS spectroscopy}

The MOS spectroscopy of both M81 fields were acquired classically on January 24, 2012, with GMOS-N at Mauna Kea. We employed both the R400 and B600 gratings in order to obtain the full spectral domain between 3200 and 9000 \AA~ needed for the science goals, and used 1\arcsec~ slitlets with 2$\times$2 binning. The structure of our observing sequence is the one suggested by Gemini, which for each field consists of: (1) the baseline arcs in both gratings, (2) the mask image, (3) the acquisition of the field, (4) the science observations with flats, with three MOS spectra for each setting. Then we also acquired twilight flats, long slit spectra of  a standard star and the calibration arcs and flats for the standard star. 

In order to avoid the spectroscopic gaps (roughly 37 pixels each) between the three detectors we have observed the fields with  B600 MOS and central wavelengths 5200 and 5250 \AA, while only one wavelength 
setting, 7400 \AA, has been used for the R400 gratings. We also acquired R400 spectra with central wavelength 5200 \AA, which we did not use in the science presented here (the overlap with he other spectra was too limited).

The data analysis has been performed with the IRAF\footnote{IRAF is distributed by the National Optical Astronomy Observatories, which are operated by the Association of Universities for Research in Astronomy, Inc., under cooperative agreement with the National Science Foundation.}  routines for GMOS. First, we have selected biases and run {\it gbias} to obtain a combined bias frame. 
For each field, grating, and setting we have then combined and reduced the flat frames with {\it gsflat}, then run {\it sgcut}, with input the combined flats, to obtain the edges of the slits. 

In order to simplify data reduction and file keeping, we produced input and output file lists for each field, grating, and setting, so we can perform data analysis on the lists rather than individual files. The first data analysis step is to run  {\it gsreduce}. We used as input the corresponding lists of spectroscopic images relative to each field and setting, and used the bias and combined flats produced as indicated above. Cosmic ray rejection is possible due to the multiple MOS images for each setting.

The arc spectra have been reduced separately for each setting, also with {\it gsreduce}, with the same bias obtained above. Arc reduction do not include flats since they had been observed in sequence with the science images, as typical for GMOS observations. Then for each configuration we chose the best arc image, and used it to find the wavelength solution with {\it gswavelength}, making certain that we used the updated line data as input while running the routine. Reduced arcs should be then transformed with {\it gstransform} to obtain the {\it wavtran} parameters. We have inspected all transformed arcs for spurious lines, taking care that the set we have used in the final calibrations are all free from problems. 

The wavelength calibration was then performed for all reduced science frames with the routine {\it gstransform}, once for each setting, using as input all reduced MOS frames relative to that setting, and for {\it wavtran} the one relative to the setting as well. At this stage, we combine all frames relative to the same field, grating, and central wavelength with {\it gemcombine}, which will preserve their multi-slit format. The end products are 4 reduced, calibrated, combined spectral frames for each field (two for each grating, one for each central wavelength). Each frame contains 27 (field~1) or 29 (field~2) spectra of the HII regions. We inspected all spectra, and verified that the major expected emission lines, such as H$\alpha$ and H$\beta$, were at the correct wavelengths. 

In order to extract the 1D spectra from each frame we used the {\it gsextract} routine, which also performs sky subtraction in the process. The manuals and instructions  available for these gemini IRAF routines are extremely scarce, thus we have used instead Massey's IRAF {\it apall} manual to determine the best ways to perform the spectral extraction. To assess the quality of our spectral data we had to obtain a sensitive function for each grating, so we could compare the fluxes of same-target lines from different gratings, where present, and also estimate Balmer line flux ratios. The sensitive functions for each grating have been derived from standard star observations.

The standard star G191B2B has been observed in longslit mode to provide flux calibration. While we do not expect excellent absolute fluxes from this project, we do need relative calibration across gratings that will provide a correct reddening and abundance analysis. The standard star observations were obtained with gratings B600 (central wavelengths 5200, 4200, 6200 \AA) and R400 (7400, 4200, 6200, 9000 \AA), to have adequate wavelength coverage and overlaps.  We retrieved the calibration arcs and flats relative to these settings  from the Gemini Data Archive. The 1D spectra were treated similarly to the 2D spectra, by (i) combining and reducing the flats, (ii) reducing the science frames, (iii) reducing the arcs, (iv) finding the wavelength solution and parameters for the arcs, and (iv) finding the wavelength solution for the science frames. 

At this stage, we found a sensitivity function solution for the standard star. First, we sky-subtracted all standard star science frames that have been previously reduced, and transformed them according to their corresponding {\it wavtran} solutions, with {\it gskysub}. Following, we proceed to extract the 1D spectra for all settings with {\it gsextract}. The next step is to obtain the sensitivity function with {\it gsstandards}, which are used to calibrate the standard stars spectra.  Give the setting of the standard star spectra, we have redundant information for the sensitivity function. We thus plotted all sensitivity functions for each grating, and obtained a spliced function per grating that covers the wavelength domain that we need to calibrate the MOS science frames. We use the same sensitivity functions for both fields.

The extracted science spectra have been calibrated with {\it gscalibrate} and the appropriate sensitivity function. We had compared H$\beta$ in both central wavelength settings of the B600 spectra for region 1, and the fluxes are within 4$\%$ in field 1, and within 0.4$\%$ in field 2. The routine {\it scombine} had been used to combine pairs of spectra with different central wavelengths, and to isolate the individual slitlets. 

In Figure 2 we show the of 1D spectra of region 1F1 and 21F1, to show the typical S/N achieved in the bright emission lines.

\subsection{Spectral analysis}

In Table 2 we list the ID (column 1), and coordinates (columns 2 and 3) of the HII regions observed. We have searched the literature for possible earlier identification of the regions, and we found possible pre-identification (within 5 $\arcsec$ of the Gemini coordinates) for several regions; we give the possible aliases in column 4.  In this table, and in the rest of the paper, we will use the target nomenclature {\it RFM}, with R=region number, and M=field number, for all HII regions analyzed. 
A {\it N} in column (5) in Table 2 means that the spectrum is not well characterized, thus we do not publish its line fluxes in the tables (either the S/N was below 3 for the whole spectrum, or a notable mismatch between the red and blue spectra was noted). Furthermore, a {\it N}  in column (6) means that weak-line abundances are not available, while a {\it U} in the same column indicates that abundances have been calculated in conditions of very low S/N plasma diagnostic lines for the temperatures, or if some of the major ionic lines were missing, thus the derived total abundances were unreliable (we still give the plasma diagnostic and ionic abundances in the tables, but do not use them for gradients or average abundances). 

We measured the emission-line fluxes with the IRAF routine {\it splot}. Errors in the fluxes had been estimated with the {\it splot} analysis, as described in detail in the {\it splot} help file. Line errors are based on the observed parameters measured in each spectrum, such as pixel count in the continuum and inverse gain, and require a model for the pixel sigmas, which is based on Poisson statistics. For blended lines, {\it splot} offers a Montecarlo modeling of the line blend. We chose N=50 iterations for the de-blending functions. All measured line intensities have been corrected for extinction when emission lines of the Balmer series were available (see notes in Table 2). In the few cases where the extinction constant is negative we assume null extinction. We have used the extinction correction formula with T$_{\rm e}$=10,000 K for all targets, then obtain, when possible, electron temperature and density, and then recalculate the extinction constant with the appropriate theoretical H$\alpha$/H$\beta$ fluxes given the target electron temperature (see Osterbrock \& Ferland 2006). Extinction-corrected line intensities are given in Table 3 (published online), where for each target we give its name, logarithmic extinction constant with its error, and then the line ID,  wavelength, corresponding  flux and its error, and the line intensity. All fluxes and intensities have been scaled in the usual way, where F$_{\rm H\beta}$=100, and I$_{\rm H\beta}$=100.

We detected lines in common in the B600 and R400 gratings for 34 regions, thus we could assess the data analysis across gratings. We calculate, for H$\alpha$, F$_{\rm B600}$=1.033$\times$F$_{\rm R400}$ - 3.59$\times$10$^{-19}$, with correlation coefficient (in the log F) of 0.96. In Figure 3 we show the residuals on the H$\alpha$ fluxes between the two gratings as a function of flux and distance from the galactic center. There is a net correlation of the residuals with flux, as expected given that fainter lines have larger relative errors. The mean relative differences between H$\alpha$ fluxes in the two gratings is 0.2$\pm$0.18; only for regions 4F1, 5F1, 7F1, and 16F1 the mean relative difference of the flux across grating is notable; such a difference is readily explained in 4F1 and 7F1, where the H$\alpha$ emission saturates the slit, and thus it is hard to get a good sky subtraction for this bright emission line. For these two targets the [S~II] doublet at 6717-31 \AA~ is present in both gratings, and we thus use these lines for the grating comparison, finding an average difference of $\sim$0.3$\pm$0.05 between the sulphur lines from the B600 and R400 spectra both for 4F1 and 7F1, making us confident that the spectral data reduction has been accurate. Regions 5F1 and 16F1 present a relative difference of H$\alpha$ across the gratings of $\sim$0.5 with no apparent reason, excluding the fact that these regions are very faint. The H$\alpha$ residuals do not correlate statistically with the distance from the galactic center (we have excluded exclude 4F1 and 7F1 from the statistical calculation, see above). 

The selection of our targets is based on H$\alpha$ pre-imaging, thus some contamination of the HII sample with other sources is possible. We are particularly concerned about PN contamination, which would affect the presumed age of the population. The spectral analysis allow us to determine the likelihood that the regions observed are indeed young  HII regions by comparing their fluxes with classical diagnostic diagrams (e.~g., Baldwin et al. 1981, Kniazev et al. 2008). In Figure 4 we show the log(I$_{\lambda5007}$/I$_{H\beta}$) vs. log(I$_{\lambda6584}$/I$_{H\alpha}$) plot for the analyzed regions, similarly to the upper panel of Fig. 3 in Kniazev et al. 2008. We note that all the analyzed regions fall well below the 
I$_{\lambda5007}$/I$_{H\beta}$ $>$ (0.61$\times$(I$_{\lambda6584}$-0.47))+1.19 curve, with the exception of one region (19F1), which might be a PN instead.  We determine the 19F1 oxygen abundance, but we exclude it from gradient and average calculations. 

It is worth mentioning that the position of regions 2F1 and 27F2 coincide, within 5$\arcsec$, with previously classified {\it possible PNe}, while 18F2 was previously classified as a PN. We confirm that 2F1 and 27F2 are bona-fide HII regions. The nature of 18F2 is still unclear because we do not have a reliable spectrum. The possible misclassification of this target does not affect the conclusions of this paper since we do not derive its abundances and thus we do not include it in the analysis. 
Diagnostic line intensities have been used to calculate electron densities and temperature from the weak-line method, as in previous works (e.g., Stanghellini et al. 2010, Magrini et al. 2009b) with the IRAF {\it nebular} package. We used either [N~II] $\lambda$5755 \AA~or [O~III] $\lambda$4363 \AA~ as the weak  temperature diagnostic line, which were available for 13 regions once we exclude those whose  diagnostic line fluxes  were below S/N$\sim$3. Electron densities were inferred from the [S~II] $\lambda\lambda$6717-6731 \AA~ doublet. When the sulfur lines were unavailable, or when  the doublet line ratio leads off the domain of the density-intensity ratio curve, towards low-densities, we assumed Ne=100 cm$^{-3}$. 

In Table 4 (published online) we give the output of the plasma and abundance analysis, including the upper limits. We give the region name, then for each region we list the parameter (column 1) and its value (column 2). Parameters include electron density and temperatures, ionic abundances of ions whose emission lines had sufficient S/N, the Ionization Correction Factors (ICFs) used in the calculation of elemental abundances, and the elemental abundances. The method used for abundance calculation is similar to that used in other papers of this series (Magrini et al. 2009b, Stanghellini et al. 2010),  where the ionic abundances are formalized in the IRAF routines {\it ionic} within {\it nebular}, and the ICF scheme is based on Kingsburgh \& Barlow (1994).

Uncertainties in the abundances are due to (1) uncertainties in the abundance diagnostic lines, which are very small being the lines very strong; (2) uncertainties in the diagnostic lines for electron temperature and density, the first being relatively high due to weak line diagnostics;  (3) uncertainties in the ICF that we have assumed for the determination of atomic abundances. 
To get errors in all ionic abundances we have propagated the errors in the diagnostic lines used in plasma diagnostics. When calculating the final atomic abundances, we need to consider both the ionic abundances uncertainties and those from the ICFs (Kingsburgh \& Barlow 1994). The oxygen ICF is equal to unity for all our regions, since ICF(O)=((He$^+$+He$^{2+}$)/He$^+$)$^{2/3}$, and we do not detect the He$^{2+}$ emission in our targets, then the uncertainty in the ICF assumption for oxygen is zero.  In the case of argon, ICF(Ar)=1.873$\pm$0.41 for all regions, thus we add the ICF error (in quadrature) to the ionic abundance uncertainties.  To obtain the final uncertainties of sulfur and nitrogen abundances we add, in quadrature, the uncertainties in the emission lines, both the ones of the abundance diagnostics and the weak-line for plasma diagnostics, to the uncertainties in the oxygen abundances. We do so since the ICFs for sulfur and nitrogen have a direct dependency on the oxygen abundances. The final uncertainties are very conservative. In Table 5 (columns 3 through 7) we give the abundances and their errors for the regions where the weak-line diagnostics was determined from diagnostic lines with S/N $>$3.

We could simultaneously calculate electron temperatures from both the low- and high-excitation emission lines only in 3 regions (1F1, 21F1, and 8F2), and we have used the ionization-appropriate diagnostics for the relevant ionic abundances. For all other regions we used the one temperature available for all ions. This translates into an additional uncertainty in the abundances, estimated to be much lower than other uncertainty factors. If we assume that the two temperatures are within 70$\%$ of one another, then the maximum error in abundance determination is 0.01 dex for the high-excitation, and 0.02 dex for the low-excitation ions.

We found almost no correlation between oxygen and argon, or oxygen and sulfur abundance sets, similarly to what we have seen in the MMT sample of HII regions in M81 (Stanghellini et al. 2010).

\section{Radial metallicity gradients, and elemental enrichment}

In this section we discuss the data points that constrain the metallicity gradient of HII regions and PNe, based on weak-line abundances of emission-line targets. In Table 5 we give the abundances used for the gradient determination.  In order to study the radial metallicity gradients  we calculate the galactocentric distances (R$_{\rm G}$) of all regions, with the method described in Stanghellini et al. (2010), and give them in Table 5 (column 2).  Uncertainties in the galactocentric distances are
mostly due to the galaxy inclination on the plane of the sky and given the moderate inclination angle of the disk and the large distance to M81 they are very low. 

Together with the HII region sample observed with GMOS, we also used the sample from MMT spectroscopy (Stanghellini et al. 2010). Of the 19 PNe whose abundances have been calculated by Stanghellini et al. (2010), 7 are observed in the GMOS fields. We confirm that 5 of these are indeed PNe, while PN~45 and PN~70 turned out to be extended;  at the distance of M81, planetary nebulae are point sources, thus we re-classify these two extended sources as HII regions and include them in the MMT HII region sample presented here, while at the same time we remove them from the PN sample. PN33, also in the PN sample where MMT and GMOS observations overlap, is very close to another HII region, thus it is hard to determine whether it is extended. To calculate radial metallicity gradients we use the HII regions in Table 5, and the HII regions and PNe in Table 5 of Stanghellini et al. (2010), but with PN~45 and PN~70 therein moved to the HII region sample. Patterson et al. (2012) also analyzed HII regions in M81, and derived a handful of weak-line abundances, moistly relevant to the large galactocentric distances, which we discuss in the next section.

The best way to calculate fits for the gradients is with the {\it fitexy} routine in {\it Numerical Recipes} (Press et al. 1988).  The {\it fitexy} routine handles both abundances and distances uncertainties, and gives not only the slope and intercept but also the likelihood of the fit, in terms of $\chi^2$ and its probability {\it q}\footnote{q gives the probability that a correct model would give a value equal or larger than the observed Chi-squared.}. Simple linear fits with least square routines can be misleading, especially if the x and y uncertainties have very different scales; in this case, the abundances are logarithmic quantities, and have variable uncertainties, while the distances have very low uncertainties in the linear scale. A discussion of {\it fitexy}  as being the best choice for astrophysical linear fits is found in Park et al (2012). Oxygen abundances are the best probes for metallicity gradients in HII regions:  They are determined without  ICF corrections, thus their uncertainties are limited to those in the line intensities, which are small with the exception of the auroral diagnostic lines for the T$_{\rm e}$ determination. We deem that gradients determined from the GMOS and MMT samples analyzed jointly are best because based on a larger  database. Both the MMT and GMOS samples alone are too small for statistical significance (the present work was strongly motivated by the limited number of HII regions observed with MMT). 

In Figure 5 we plot the oxygen abundances vs. galactocentric distance for different populations. The top panel of Figure 5 shows HII regions from both the MMT and GMOS samples; for this joint sample we found -- by setting distance uncertainties to 0 -- a negative oxygen gradient $\Delta$log(O/H)/$\Delta$R$_{\rm G}$=-0.088$\pm$0.013 dex kpc$^{-1}$, with intercept 9.202$\pm$0.107. The fitexy routines gives q=0.051 for this gradient, which is acceptable ($\chi^2$/(degrees of freedom) $\sim$1). The data have a lot of scatter (see Fig. 5), thus we did not expect a very high probability. We also determine that the fit residuals are spread between 0 and 0.7 dex, peaked at 0.2, where the abundance errors are typically between 0.1 and 0.5 dex.  It is worth noting that the MMT sample alone would not produce a reliable gradient, given the limitedness of the sample, as assessed in Stanghellini et al. (2010). The resulting oxygen gradient is consistent within the error with the  strong-line gradient by  Garnett \& Shields (1987). 

For the newly defined planetary nebula set (lower panel of Fig. 5, also in Table 5 of Stanghellini et al. 2010, minus PN45 and PN70), we found $\Delta$log(O/H)/$\Delta$R$_{\rm G}$=-0.044$\pm$0.007 dex kpc$^{-1}$, and intercept 8.587$\pm$0.062.  Residuals are slightly smaller for PNe than for HII regions, and they are consistent with the abundance errors; the fit probability, q=0.053, is better than, but comparable to, that of the HII regions fit. The two samples, HII regions and PNe, are thus directly comparable.  We infer a moderate evolution of the radial metallicity gradient with time. In Figure 6 we plot the top and bottom panels of Figure 5 together, to view the different gradients. While HII regions mark the current galactic time, thus their $\alpha$-elemental abundances narrate the evolutionary history of the galaxy since its formation, PN $\alpha$-element abundances capture the metallicity makeup at progenitor formation. We did not observe Type I PNe in M81; since Type I PNe have higher mass progenitors (Peimbert \& Torres-Peimbert 1983), we can assume that all PN progenitors in M81 are roughly in the 1-2 M$_{\odot}$ mass range, with a population age between 1 and 10 Gyr  (Maraston 2005, see also Stanghellini et al. 2010). 

Oxygen abundances and the location of the probes seem to indicate that the radial gradient determined by HII regions is steeper than that of PNe (see Table 6). It is worth noting that the radial range of PNe is more extended that that of HII regions. If we rigorously select only PNe in the radial range defined by the HII regions we would have only 8 targets left for the gradient estimate. Such small sample is not adequate to get a gradient with the {\it fitexy} routine. A simple least-square fit to these 8 data points gives an even flatter gradient, -0.03 dex kpc$^{-1}$, which is again compatible with the results above. It is worth noting that several HII region abundances carry very large uncertainties. If we select only HII regions with $\Delta log O/H<0.3$ (middle panel of figure 5) we find a more constrained set of regions, where a direct least square fit gives a slope of -0.07 dex kpc$^{-2}$.  Finally, if we were to eliminate from the sample of HII regions the one with smallest galactocentric radius we would obtain a similar fit with slope -0.08$\pm$0.015 dex kpc$^{-1}$ and q=0.05; if instead we were to eliminate the HII region with largest galactocentric distance, the {\it fitexy} routine would not converge. In summary, while statistical analysis indicates that the radial metallicity gradient is steepening with time,  clearly  more abundances of regions with larger galactocentric distances are needed to pin down the gradient evolution for this galaxy. This has implications in our conclusions (see Section 5). 

The indication of a gradient evolution that we found covers a rather limited galactic baseline, and it can not be extended to $\alpha$-elements other than oxygen, given the limited samples for which neon and argon abundances are available in HII regions and PNe of M81, and given the intrinsic uncertainties of sulfur abundances. If we include the four HII regions with weak-line oxygen abundances and with R$<$16 kpc (i.e., the inner regions) in Patterson et al. (2012) in the sample we would obtain a gradient slope of $\sim$-0.08$\pm$0.01 dex kpc$^{-1}$, and an average oxygen abundance of 3.4$\pm$1.9$\times$10$^{-4}$, thus none of the conclusions regarding gradient evolution or enrichment  would change notably. The outer regions analyzed by Patterson et al. (2012) are discussed in the next section.

Nitrogen also indicates the presence of a similarly negative gradient slope. Figure 7 shows the nitrogen gradient for HII regions and the PNe, similarly to Figure 5.  Nitrogen is not an $\alpha$-element, thus gradient evolution is the result of both enrichment and AGB evolution, thus the slopes of HII regions and PN gradients could not be applied to infer the (lack of) gradient evolution. It is worth noting that the fit of nitrogen gradients for the HII regions presented in Figure 7 does not converge to a suitably low $\chi^2$ value; both the complete HII region sample and the more limited low-error sample fits produce rather high $\chi^2$ values with $\chi^2$/(degrees of freedom) $\sim$10, with low fit probabilities. In Table 6 and in the figure we give the least-square fits, weighted with the uncertainties, which does not produce fit uncertainties.

Although sulfur abundances are notoriously unreliable, we still report them in the abundance tables but we do not feel they should be used for detailed gradient analysis.  The argon gradient of the combined sample is also moderately negative, but with a relatively larger error bar. It is worth noting that the ICF for the sulfur abundances is calibrated on PNe (Kingsburgh \& Barlow 1994), while different calibrations seem to fit HII regions better, and in general sulfur abundances suffer from unpredicted deviations from expectations.  A  reasonable sample of PN for argon gradient determination is not available.

We give in Table 6 the abundance means, in logarithmic form, with their uncertainties.  It is worth noting that all elements present a high level of scatter in the M81 sample, and that the HII region abundances are more scattered than the PNe. A measure of an average galactic elemental enrichment can be inferred from the average abundances of HII and those of the PNe. Enrichment is noted in all $\alpha$-elements studied here. Since we are comparing HII regions and non-Type I PNe we are probing a time lag of about 6 Gyr, on average, in stellar populations (see also Maraston 2005; Stanghellini et al. 2010). The amount of $\alpha$-elemental enrichment is 0.14$\pm$0.08 dex for oxygen, which is consistent with marginal enrichment. Naturally, given the large scatter of the HII region abundances, such enrichments are only an indication of a trend. Enrichment values calculated for the other $\alpha$-element are similar to that of oxygen, but the high scatter and the small sample sizes make the uncertainties high and the result not statistically significant.

\section{Discussion}
\subsection{A break in the radial metallicity distribution?}

In Figure 8 we show our HII region oxygen abundances vs. distances plotted together with those HII regions analyzed by Patterson et al. (2012) whose abundances are based on weak-line plasma diagnostics, thus directly comparable to our sample. The broken line, based on the oxygen gradient  given in Table 6, is compatible with the inner regions from Patterson et al. (2012). The region at 16 kpc is  
M\"unch-1, whose abundance of log(O/H)= 8.1 is compatible, within the errors, with the inner gradient, and it is $\sim$0.4 dex lower than the average abundance of the inner regions. On the other hand, region 28 (or KDG~61, farther than 30 kpc from the center) is incompatible with the inner gradient, and indicates gradient flattening in the outer regions of M81. The HII region coincident with the position of the dwarf galaxy KDG~61 has been studied in detail by Makarova et al. (2010). The  redshift difference between the HII region and the galactic stellar population is higher than the expected velocity dispersion within the dwarf galaxy, thus the HII region most likely belongs to the M81 disk.  On the other hand, there is some uncertainty on whether M\"unch-1 really belongs to the M81 disk. 

Based on the best data available, those in Figure 8, the resulting scenario is compatible with a negative gradient in the inner disk and a basically flat gradient in the outer disk, but in order to confirm the sharp abundance break between approximately 10 and 16 kpc (R$_{\rm 25}$=14.6 kpc in M81, Scarano \& Lepine 2013) one would need to collect additional data in the general distance range where M\"unch-1 is located. The field surrounding M\"unch-1 is very rich in HII regions (Greenawalt et al. 1998), whose spectroscopic analysis is feasible with 8m class telescopes. It is worth recalling that breaks in the radial metalicity gradients have been already observed in several spiral galaxies.  Bresolin et al. (2009, 2012) have shown that a break around R$_{\rm 25}$ is evident in several large spirals such as NGC3621 and NGC1512; unfortunately, these conclusions are based for the most part on strong-line abundances, corroborated only by 3-4 weak-line data points per galaxy. While the metallicity breaks seem evident in these galaxies, the strong-line abundances do not always agree with the weak-line abundances, thus these results remain to be confirmed. 

From the modeling viewpoint, a break in the radial metallicity distribution of a spiral galaxy could be produced by (1) mixing and turbulence processes (e.g. effects of radial gas flows induced by bars); (2) galactic scale outflows, which would explain the enrichment of the circumgalactic and intergalactic medium via galactic winds (e.g., Tumlinson et al. 2011), and the origin
of the mass -- metallicity relation (Finlator \& Dav{\'e} 2008); and  (3)  enriched accretion, whose importance in defining the observed galaxy mass -- metallicity and mass -- gas fraction relations at z$>$ 1 has been underlined by Dav{\'e} et al. (2011).

\subsection{Comparison of M81 to other galaxies}

Stanghellini \& Haywood (2010) have shown that there is indication that the radial metallicity gradients are steeper for the young than the old stellar populations in the Galaxy. In M33 the young and old populations have marginally different gradient slopes, and the gradient evolution is minor (Magrini 2009b, 2010). Both results have been derived from emission line objects through weak-line abundance analysis.
In this paper we have indicated that the oxygen gradients for PNe and HII regions are negative, and that they are different within the errors. We have also shown that the log(O/H) vs.  R$_{\rm G}$ distributions for HII regions and PNe are statistically different, and consistent with the radial oxygen gradient of HII regions steeper than that of PNe.
We illustrate this in Figure 9, where we plot the oxygen gradient slope, in dex kpc$^{-1}$, against the average age of the probing stellar population. We place the HII regions at t=0, and the PN population at their likely average age, as described in Stanghellini et al. (2010), considering that there are no Type I PNe in M81. The age error bar covers the age-span of the PN population. The plot seems to indicate that there is in fact an evolution of the radial metallicity gradient in M81. This conclusion holds when including Patterson et al. 's (2012) data for the inner HII regions.

We like to add to the plot of Figure 9 {\it all galaxies whose metalicity gradients have been studied with weak-line oxygen abundances of PNe and HII regions.} It would not be straightforward to add Sanders et al.'s (2012) data for M31 to Figure 9. In fact, only 4 HII regions in M31 have weak-line abundances, thus a gradient based on them would be very uncertain. The PNe in M31, according to Sanders et al. (2012), show a flat weak-line gradient.  The only other spiral galaxy that to our knowledge has enough PN and HII region abundances from weak-line analysis to be compared with the others of Fig. 9 is NGC 300 (Stasinska et al. 2013). In NGC 300 the PN gradient is shallower than the HII region one. If we translate the gradients given by Stasinska et al. (2013), which are referred to R$_{\rm 25}$, to gradients in dex kpc$^{-1}$, we obtain $\Delta$log(O/H)/$\Delta {\rm R_G} $=-0.024$\pm$0.014 (PNe), and -0.068$\pm$0.0091 (HII regions), as shown in Fig. 9. 

While the oxygen gradient slopes plotted in Figure 9 are all based on weak-line abundances, and they seem to indicate that the metallicity gradients for the PNe are always flatter than those of HII regions, we are well aware that each of these derivations have its own selection effects and intrinsic problems. For the Galactic sample,  the distance uncertainty problem is present, although greatly alleviated by the use of the Magellanic Cloud distance calibration (Stanghellini et al. 2008). Abundances of PNe and HII regions are based on temperatures derived from auroral lines of [O III] or [N II];  it is well known that emission from the O$^{+2}$ ion acts as major coolant for HII regions and PNe, whose efficiency tends to increase with metallicity, making the corresponding auroral emission line very faint;  a selection against metal-rich regions is thus possible when based on this emission line. Our regions in M81 have been selected through H$\alpha$ imagery, thus such a selection effect is minimal in our sample. A marginal selection effect toward high H$\alpha$-emission regions is possible. 

Planetary nebula models that take into consideration the  oxygen abundance of the population show that the intensity ratio of the [O III] 5007\AA~ emission line vs. H$\beta$ declines with metallicity if we select ionizing stars at the peak of their temperature evolution (Stanghellini et al. 2003). By running Cloudy (Ferland et al. 2013) models similar to those of Stanghellini et al. (2003) we find that the 4363 \AA~intensity relative to H$\beta$ decreases from $\sim$0.2 to $\sim$0.1 going from solar to SMC metallicity, all other input parameters being identical. A similar situation probably occurs for HII regions, and the reason is that metal-poor environments favor the production of carbon-rich stars. This effect has the consequence that  HII regions in low-metallicity environments have higher carbon abundances; consequently, their cooling is provided mostly by the CIII line in the UV. While UV observations should help in quantify this behavior it would seem improbable that a huge selection effect against high metallicities is at place in M81. 

The results of Figure 9 are compatible with steepening of the metallicity gradient with time, with the spiral galaxies forming inside-out, with infall of pre-enriched gas, or enhanced feedback. Gibson et al. (2013) have produced metallicity gradients in a cosmological context by modeling isolated, Milky Way-type galaxies at various redshifts and assuming a priori that the formation is inside-out. For the isolated disk galaxy models with enhanced feedback they reproduce a curve that fits perfectly the Milky Way data in Figure 9 if the look back ages were translated into redshift with a cosmological calculator (Wright 2006) and standard assumptions (open CDM model). Gibson et al. (2013) found, with their MaGIC g1536 simulation, that the gradients are rather flat for all populations back to $z\sim$1.5, with the (negative) slope steepening with time.  We encounter this same behavior in all spiral galaxies where both PN and HII region gradients could be determined by weak-line analysis. For M81, the comparison should be done with models that include group components.

Interestingly, Gibson et al.'s (2013) models indicate that galaxies forming inside-out in groups seem to show steeper metallicity gradients than those of galaxies in isolation. M81 is the only galaxy in the set discussed in our Fig. 9 belonging to a loose group, and it also happens to display the highest (negative) gradient slopes. The environment can affect the gradient slope;  metallicity gradients for both PNe and HII regions seem to be steeper in M81 than in the Galaxy, as noted by  Stanghellini et al. (2010).  It is worth noting that Sanchez et al. (2013) found that metallicity gradients of HII regions in spiral galaxies belonging to groups  are flatter than those in isolation. While we can not assess whether the gradient evolution that we find is compatible with Sanchez et al.'s work (they do not quote gradient uncertainties) we note that they use strong-line oxygen abundances, which can give different results from the weak-line analysis, and secondly they estimate metallicity gradients in the 0.3-2$\times$ R$_{\rm 25}$ radial domain, whether ours is limited to R$_{\rm 25}$  $<$ 1. 

Without concluding that all galaxies in Fig. 9 have had  an inside-out chemical evolution and pre-enriched gas infall, this hypothesis of galaxy formation would be  compatible with the observed framework.  While we find a mild evidence for gradient steepening with time, compatible with the scenario described by Gibson et al (2013), we are aware that inside-out galactic disk formation may also produce  gradient flattening with time, depending on other assumptions, thus our findings are by no means a clear indication that the inside-out disk formation is at work.  With all other conditions being equal, the radial metallicity gradient seem to be negatively steeper for the lower-mass galaxy (only stellar mass is considered here). 

We also took into account the Hubble type of the observed galaxies, as given by Zaritsky et al. (1994), when comparing the radial oxygen  gradients for the galaxies of Figure 9 to one another. We multiplied the gradient slopes by R$_{\rm 25}$ \footnote{the effective radii are from the NED catalog at
http://ned.ipac.caltech.edu/} for the comparison. We found that the (HII region) oxygen gradient slope relative to R$_{\rm 25}$ for M33 (type 6) and NGC~300 (type 7) are very similar, respectively -0.042 and -0.36 dex/R$_{\rm 25}$, as expected given their similar Hubble type, while the M81 (type 2) absolute gradient slope is $\sim$-1.3 dex/R$_{\rm 25}$, which seems to indicate a steeper gradient for an earlier Hubble type galaxy. This result is not in contrast with Zaritsky et al.'s (1994) trends, and may instead indicate either that the environment affects the Hubble type-gradient trend, or that the break in the M81 galactic gradient carries more weight than seen so far with the available data. 

It is worth noting that the inside-out galaxy formation associated to pre-enriched infall is not the only explanation of the presented data. In fact, models 
of galaxy evolution produce a degeneration of results if we consider radial migration. If for example we consider the models by Kubryk et al. (2013), or Michev et al. (2013), we see that the older stellar populations suffer a much larger migration than the young ones, thus the observed gradient difference between PNe and HII regions could be due in principle solely to stellar migration. Kubryk's (2013) models invoke a strong bar, which might not be the case for M81. In the Galaxy, appropriate de-migration of the observed PN populations would compensate the observed metallicity gradient evolution (Stanghellini \& Haywood, in preparation). While beyond the scope of this paper, it is worth underlying that current models can not be uniquely constrained by the available data.

\section{Conclusion, and future work}

Weak-line abundance analysis is performed for a sample of HII regions in M81, based on GMOS/Gemini multi-object spectroscopy.  Together with other datasets we have collected in the past with the MMT we found oxygen enrichment of 0.14$\pm$0.08 dex in M81.

We also found a radial metallicity gradient $\Delta$log(O/H)/$\Delta$R$_{\rm G}=-0.088\pm$0.013 dex kpc$^{-1}$, when using HII regions as probes. Compared to the PN gradient, which is recalculated here based on new PN identifications, to be -0.044$\pm$0.007 dex kpc$^{-1}$, this result is consistent with the metallicity gradient steepening with time since galaxy formation, if stellar migration is not accounted for. Compared to other galaxies for which these diagnostics are available, there is consistency of gradient steepening with time in the Galaxy, M33, and NGC 300, in addition to M81. It appears that 
the gradient has steepened in M81 more than in the other galaxies examined by weak-line abundances.

Our M81 data are consistent with a negative HII region oxygen gradient in the inner galaxy, but they can not exclude a flat gradient in the outer galactic regions, as indicated by Patterson et al.'s (2012). A better handle on this possibility could be offered by observing other M81 fields in the outer zones where the gradient seems to break. This would actually be the first direct test of a radial metallicity break by using only weak-line abundances.

The results obtained in this work are based solely on weak-line abundances, from direct empirical methods. We avoid to mix the weak-line and strong-line derived abundances in order to obtain as pure a sample as possible. By deriving abundance using the strong-line method, possibly based on calibrations within our own observations, would have certainly enlarged the sample size and lower the formal gradient errors. Abundances derived from the two methods have different errors and might have different systematics.  Furthermore, abundances from strong-line fitting formulae can have intrinsic errors of $\sim$0.5 dex, and the gradients in spiral galaxies are very shallow in general, which makes a bad combination for precise gradient studies.

Both the abundances presented here for M81 from GMOS, and those from the MMT observations, have large error bars. We also note that, while statistically accurate, the difference in gradient is below 3$\sigma$ and can be seen as tentative at this stage. Although we have been conservative in the error bar estimates, it is clear that this abundance determination method has been pushed very far here. It is worth noting that the GMOS time allocated for this project was about 70$\%$ of the time requested, thus the S/N ratio for some of the diagnostics lines of regions that are listed now as lower limits to fluxes could have been completely analyzed with the original allocation of time. We plan in the future to observe additional HII regions to better define the metallicity gradient both for the inner and outer regions of this galaxy, in particular to augment the radial extent of the analysis. Also, we plan to extend this type of analysis to other spiral galaxies with different masses, metallicities, and environment conditions, to enlarge the database to study metallicity gradients and their evolution.

More progress could be taken forward with the data on hand by modeling the regions studied with photoionization analysis in a self-consistent way. We aim to reproduce the observed abundances with the observed emission lines, and to continue enriching the sample of spiral galaxies for which these diagnostics will be available.

\begin{acknowledgements}
Based on observations obtained at the Gemini Observatory, which is operated by the Association of Universities flatfieldedfor Research in Astronomy, Inc., under a cooperative agreement with the NSF on behalf of the Gemini partnership: the National Science Foundation (United States), the National Research Council (Canada), CONICYT (Chile), the Australian Research Council (Australia), Minist\'erio da Ci\^encia, Tecnologia e Inovac\~ao (Brazil) and Ministerio de Ciencia, Tecnologia e Innovacion Productiva (Argentina).  Data presented here are from Gemini programs GN-2011B-Q-32 and GN-2011B-C-3.  We thank Gemini staff for their help in supporting these observations.  This research has made use of the NASA/IPAC Extragalactic Database (NED) which is operated by the Jet Propulsion Laboratory, California Institute of Technology, under contract with the National Aeronautics and Space Administration. We are grateful to the Referee for constructive suggestions on an early version of this paper.
\end{acknowledgements}

\begin{figure}
\centering
\includegraphics[width=\hsize]{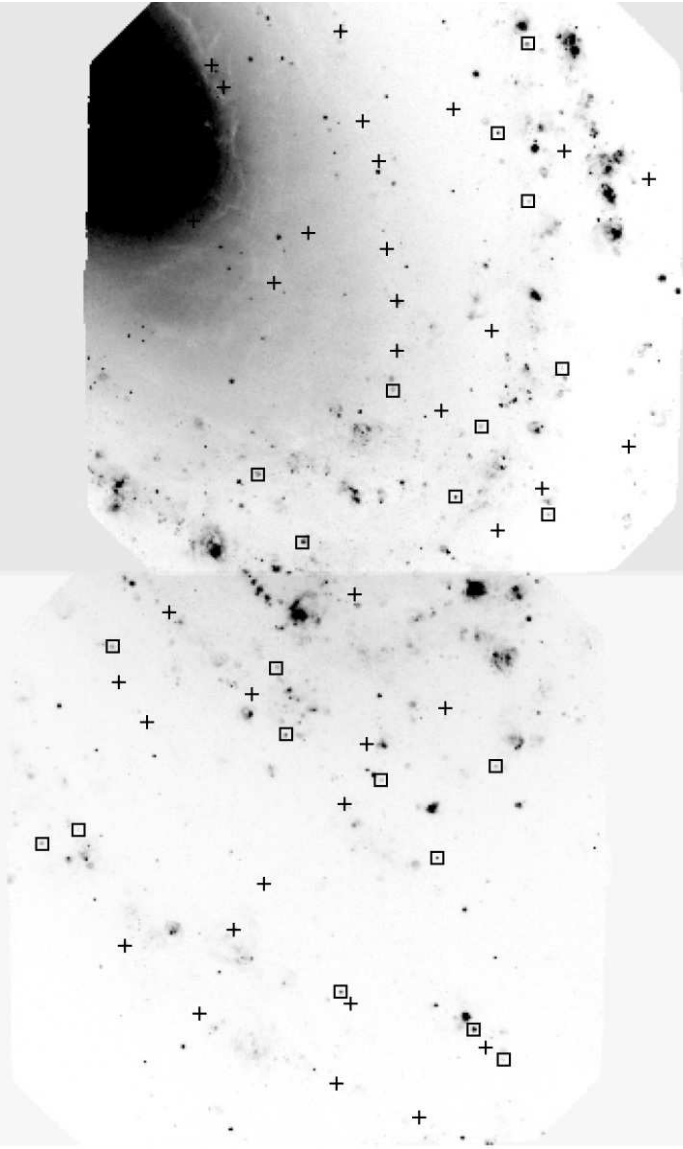}
\caption{The observed fields in the H$\alpha$ filters: field 2, the one closer to the galaxy center, is in the upper panel, while field 1 is in the lower panel. The observed regions are indicated, where the squares are the regions with measured spectroscopy. Orientation is North down, East left.}.
\end{figure}

\begin{figure}
\centering
\includegraphics[width=\hsize]{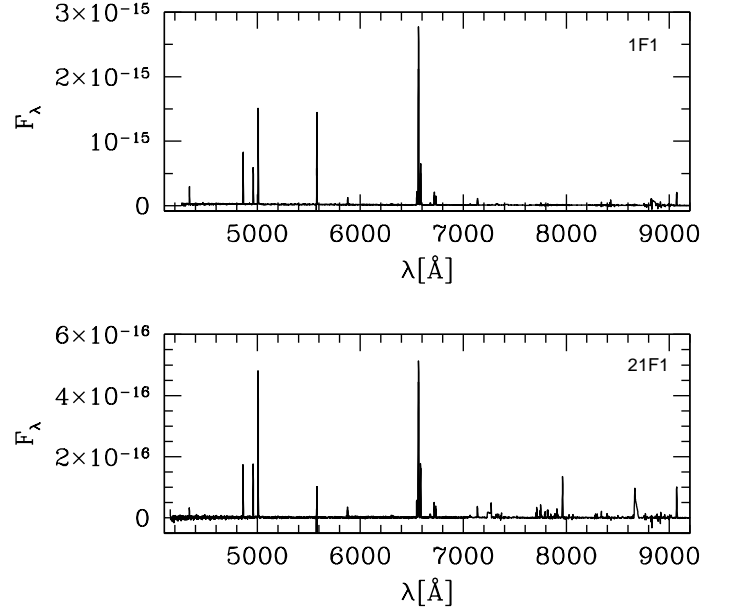}
\caption{Composed B600 and R400 spectra of 1F1 (top panel) and 21F1 (bottom panel), with fluxes in erg cm$^{-2}$ s$^{-1}$. }
\end{figure}

\begin{figure}
\centering
\includegraphics[width=\hsize]{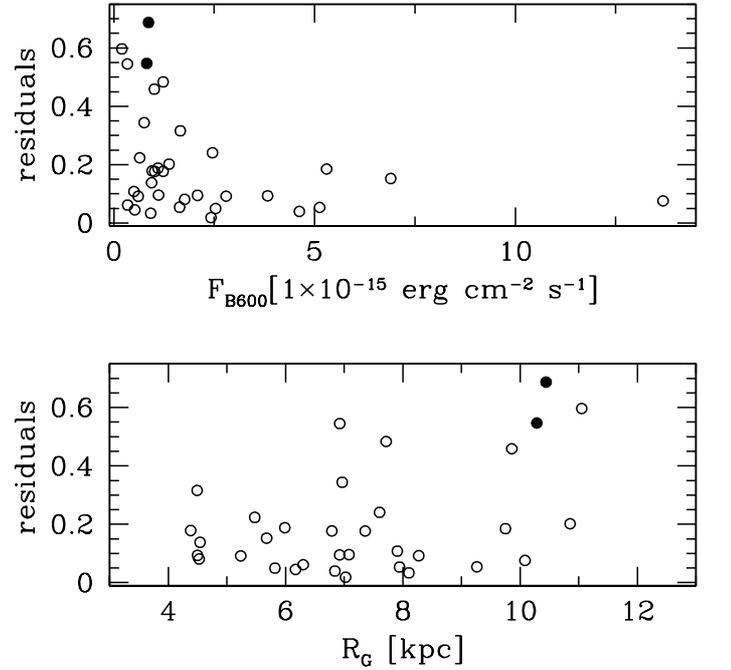}
\caption{Comparison of H$\alpha$ fluxes across gratings. Top panel: H$\alpha$~residuals ($\Delta$F/F$_{\rm B600}$) against F$_{\rm B600}$; bottom panel: residuals against the distance from the galactic center. Filled circles in both panels indicate the location of 4F1 and 7F1 (see text). }
\end{figure}

\begin{figure}
\centering
\includegraphics[width=\hsize]{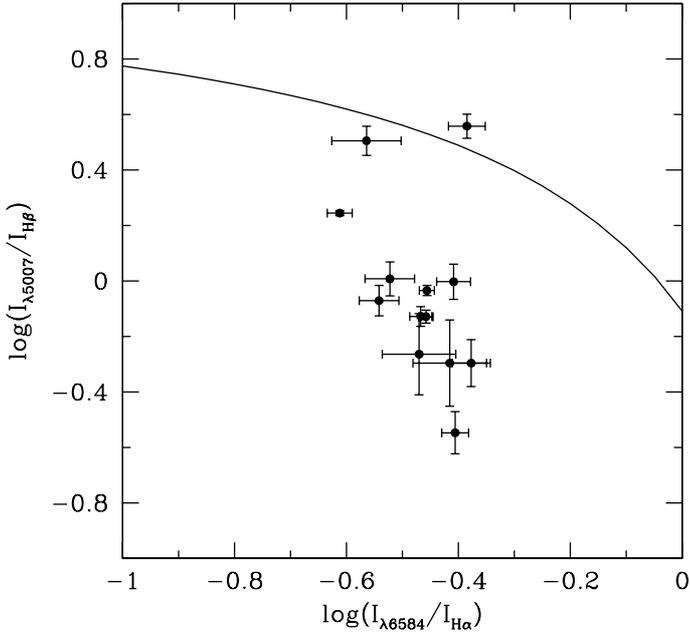}
\caption{Diagnostic plot for HII regions with abundance determinations. Regions below the curve are confirmed HII regions (see text). The region above the curve is 19F1, probably a PN.}
\end{figure}

\begin{figure}
\centering
\includegraphics[width=\hsize]{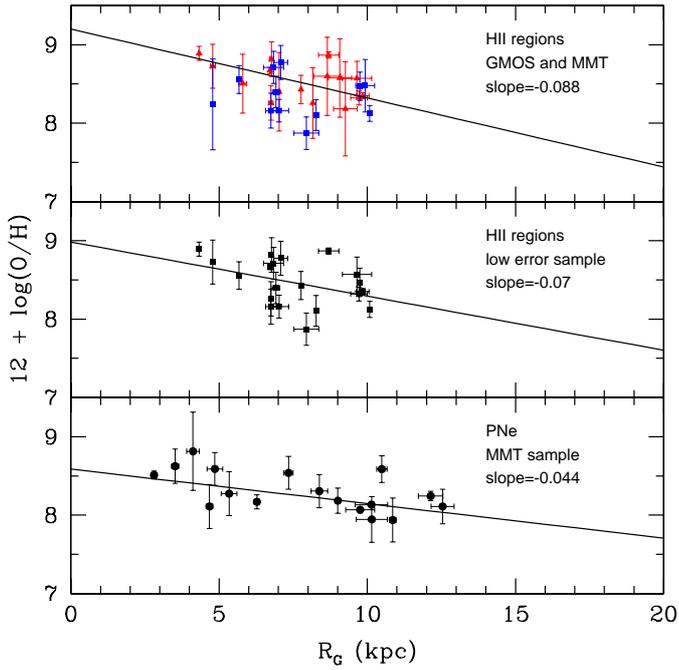}
\caption{Oxygen abundances, from weak-line analysis, versus distance from the galactic center. Top panel: HII regions from the GMOS (squares) and MMT (triangles) samples. Middle panel:  Selected sample with $\Delta log(O/H)<0.3$. Bottom panel: PNe from MMT observations. }
\end{figure}

\begin{figure}
\centering
\includegraphics[width=\hsize]{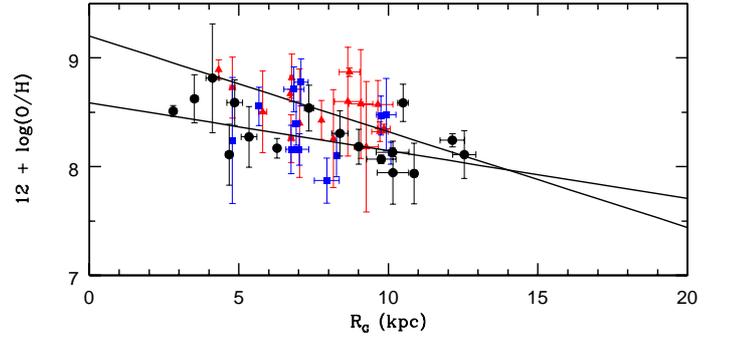}
\caption{Oxygen abundances, from weak-line analysis, versus distance from the galactic center, for HII regions and PNe. Symbols are as in Figure 5. Lines are the fits as in Figure 5 to the MMT+GMOS HII region samples, and the PNe.}
\end{figure}

\begin{figure}
\centering
\includegraphics[width=\hsize]{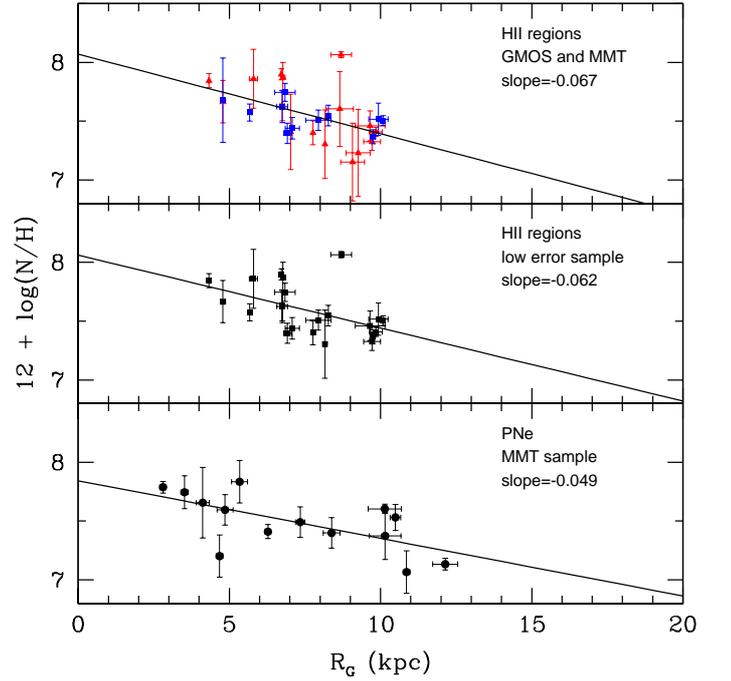}
\caption{Nitrogen abundances from weak-line analysis versus distance from the galactic center. Panels, symbols, and lines as in Fig. 5.}
\end{figure}

\begin{figure}
\centering
\includegraphics[width=\hsize]{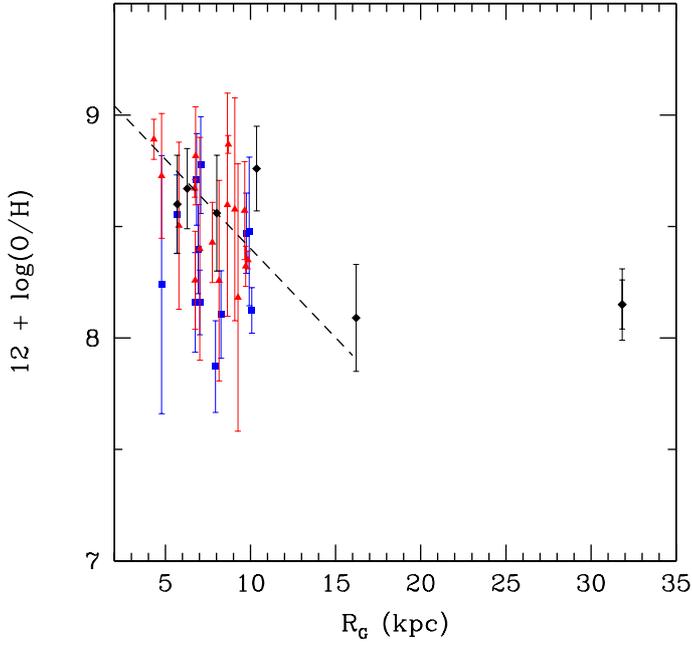}
\caption{Oxygen vs. galactocentric distance in HII regions of M81, as in Figure 5 (bottom panel), but including Patterson et al. (2012)'s weak-line abundances (diamonds). The line correspond to the gradient for R$_{\rm G}<$16 kpc, including the data from Patterson et al. (2012) with R$_{\rm G}<$16.}
\end{figure}

\begin{figure}
\centering
\includegraphics[width=\hsize]{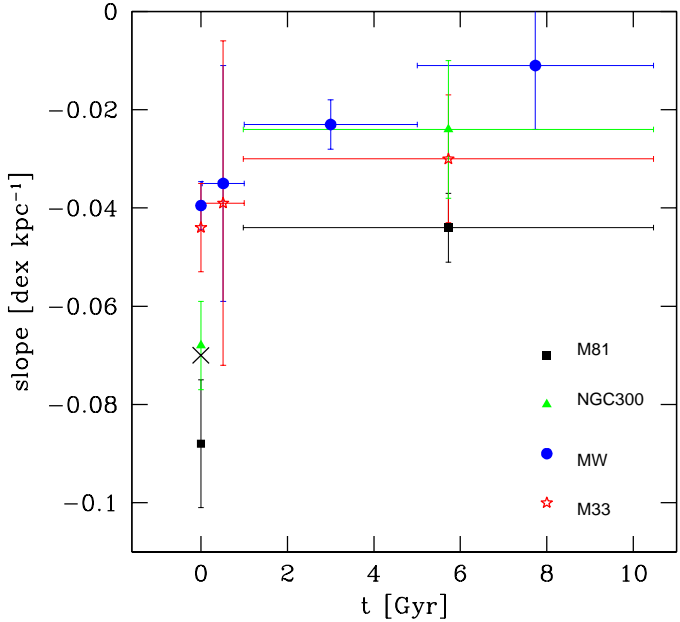}
\caption{Evolution of the oxygen gradient for several galaxies, derived from the PN and HII region gradients, adapted from Stanghellini et al. (2010). The data points for M81 have been updated considering the gradient derived here, and include all slopes of Figure 5: squares are for best statistical gradients, with converging fits and $\chi^2$ /(degrees of freedom) $\sim 1$; the cross at t=0 indicates the location of the gradient from the middle panel of Figure 5, where only low-uncertainty data points have been included. The data points for NGC 300 are from Stasinska et al. (2013).}
\end{figure}

\clearpage

\begin{table*}
\caption{Observing log}
\label{table:1}
\centering
\begin{tabular}{cclcccc}
\hline\hline
M81 field & date&mode & t$_{\rm exp}$ & IQ &
CC& BG\\
&& & [m]& $\%$& $\%$&  $\%$  \\
\hline
&&&&&&\\
{\it Pre-imaging}\\

1&	2011-12-29&	Ha~G0310& 	9&	70& 50& 50 \\
1&	2011-12-30&	HaC~G0311&	9&	85& 50& 50	 \\

2&	2011-12-08&	Ha~G0310&	9&	20& 50& 80\\
2&	2011-12-08&	HaC~G0311&	9&	 20& 50& 50\\
&&&&&&\\
{\it MOS}\\                 

1&	2012-01-24&	B600+G5307&	100&	\dots& \dots& \dots \\
1&	2012-01-24&	R400+G5305&	100&	\dots& \dots& \dots \\

2&	2012-01-24&	B600+G5307&	 100&	\dots& \dots& \dots \\
2&	2012-01-24&	R400+G5305& 100&	\dots& \dots& \dots \\
&&&&&&\\
\hline

\end{tabular}
\end{table*}

\clearpage

\begin{table*}
\scriptsize
\caption{Regions with spectroscopy}

\begin{tabular}{lccccc}
\hline\hline
 Region & RA & DEC& alias& Spectrum& Analysis\\

{\it Field 1}&&&&&\\

1F1 &      148.752594 &       69.218063  & \dots&  Y& Y\\
2F1&      148.816315 &       69.225471   & [MPC2001]  33& Y& U\\
3F1 &      148.811661 &       69.210587  & [IW2001] P32& Y& Y\\
4F1\tablefootmark{b} &      148.740646 &       69.223289 & \dots&Y& U \\
5F1 \tablefootmark{c} &      148.780655 &       69.231766& \dots&Y& N  \\
6F1 \tablefootmark{d}&      148.807785 &       69.212517 & \dots&Y& N  \\
7F1\tablefootmark{b} &      148.747757 &       69.221062& \dots&Y& N   \\
8F1&      148.876114 &       69.212425  & \dots&Y& N\\
9F1 &      148.857819 &       69.199265  & PSK 213& Y&  N   \\
10F1 &      148.907730 &       69.200500  &PSK 249&Y&  U \\
11F1&      148.842789 &       69.161339   &  \dots&Y& U  \\
12F1 &      148.804214 &       69.180344   & [PR95] 51114&Y& U \\
13F1\tablefootmark{e} &      148.942169 &       69.183174  & PSK 279&Y& Y\\
14F1 &      148.925201 &       69.181305   & \dots&Y&  N\\
15F1&      148.843521 &       69.192276   &  \dots&N &N \\
16F1\tablefootmark{c}  &      148.755325 &       69.165710   &  \dots&N &N  \\
17F1 &      148.890976 &       69.164536   &  \dots&Y& N  \\
18F1 &      148.792603 &       69.170776    &  \dots&Y& N \\
19F1\tablefootmark{f} &      148.828598 &       69.168373 & PSK 185&Y& Y\\
20F1 &      148.786636 &       69.176697   &  \dots&Y& N\\
21F1 &      148.764053 &       69.190033  & PSK 122&Y& Y\\
22F1 &      148.734604 &       69.175850  & [PR95] 51109&Y& N\\
23F1 &      148.902008 &       69.158043    & \dots&Y&  N \\
24F1 &      148.830505 &       69.157364   & PSK 186& Y& U\\
25F1 &      148.903839 &       69.152161   &PSK 246&Y& Y\\
26F1 &      148.877029 &       69.147202    & \dots&Y&  N \\
27F1 &      148.792679 &       69.146423  & [IW2001] P30&Y& U\\
&&&&&\\

{\it Field 2}&&&&&\\

1F2 &      148.703598 &       69.135574  & PSK 47&Y& Y\\
2F2 &      148.814636 &       69.137360 & PSK 171& Y& Y\\
3F2 &      148.726929 &       69.137512   & \dots& Y& N \\
4F2 &      148.832352 &       69.125946  &PSK 189& Y& U\\
5F2 &      148.730942 &       69.120667  &PSK 87& Y& N\\
6F2 &      148.664642 &       69.125381  & \dots& Y& N \\
7F2 &      148.705368 &       69.131226   & \dots&N &N\\
8F2 &      148.744461 &       69.131775  &\dots&Y& Y\\
9F2 &      148.769394 &       69.113853  & PSK 128&Y& N\\
10F2 &      148.747971 &       69.117508   &\dots&Y&  N \\
11F2 &      148.692734 &       69.111992  & \dots&Y& Y\\
12F2 &      148.766266 &       69.107254   & PSK 125&Y& N \\
13F2\tablefootmark{g} &      148.764374 &       69.099457  &PSK124&Y&  N  \\
14F2&      148.819794 &       69.095055   & \dots&Y& N \\
15F2 &      148.722977 &       69.105118   & \dots&Y& N \\
16F2&      148.853516 &       69.083992  & [PR95] 50645&N &N \\
17F2 &      148.767609 &       69.090866  & \dots&Y& N \\
18F2&      148.802048 &       69.087212   & [MPC2001] 21&N &N \\
19F2 &      148.713989 &       69.073196  &PSK 63&Y& Y\\
20F2 &      148.685257 &       69.076797   &\dots& Y& N \\
21F2 &      148.648117 &       69.082268   & \dots&Y& N \\
22F2 &      148.733612 &       69.068993   & \dots&Y& N \\
23F2 &      148.774338 &       69.069771   & \dots&N &N \\
24F2 &      148.702347 &       69.084763  &PSK 45&Y& Y\\
25F2 &      148.768829 &       69.076431   &\dots& Y& N \\
26F2 &      148.840637 &       69.059006 & [PR95] 50476&Y& N   \\
27F2\tablefootmark{g} &      148.835815 &       69.062828 &MPC2011 42&Y&  U \\
28F2 &      148.698318 &       69.059174  &[PR95] 50475&Y& Y\\
29F2 &      148.781937 &       69.055061   &[PR95] 50451&Y&  N \\
&&&&&\\

\end{tabular}

\tablefoot{
\tablefoottext{a}{[MPC2001]: Magrini et al (2001);  [IW2001]: Immler \& Wang (2001); PSK: Petit et al. (1988); [PR95]: Perelmuter \& Racine (1995).}
\tablefoottext{b}{H$\alpha$ emission is saturated;}
\tablefoottext{c}{Considerable difference between B600 and R400 fluxes ($\sim$50$\%$);}
\tablefoottext{d}{Reddening correction not available;}
\tablefoottext{e}{H$\beta$ or H$\alpha$ have a non-Gaussian shape;}
\tablefoottext{f}{Emission line diagnostics indicate that it might be a PN;}
\tablefoottext{g}{This target shows a WR-type spectrum with stellar emission lines at $\lambda$4658 and $\lambda$5812, and  C~IV features.}
}

\end{table*}

\begin{table*}

\caption{Galactocentric distances and Elemental abundances}
\begin{tabular}{lrrrrrrr}
\hline\hline
{Name}& {R$_{\rm G}$ [kpc]}& {He/H}& {log(N/H)+12}& {log(O/H)+12}& {log(S/H)+12}& {log(Ar/H)+12}\\
\hline
&&&&&&&\\
  1F1      &     10.08$\pm$  0.017 & 0.084$\pm$ 0.001      &  7.504  $\pm$0.151      &  8.124 $\pm$0.102      &  7.052  $\pm$0.146      &  6.379  $\pm$0.425     \\
    3F1      &      9.75$\pm$  0.080 &  0.094$\pm$  0.001      &  7.367  $\pm$0.261      &   8.470 $\pm$0.180      &  6.592  $\pm$0.260      &  6.198  $\pm$0.451     \\
   13F1      &  9.93$\pm$ 0.322 & 0.149$\pm$  0.004      &  7.517$\pm$0.491     &  8.478 $\pm$0.334      &  6.762  $\pm$0.475      &  6.195$\pm$0.546     \\
   21F1      & 8.271$\pm$   0.005 & 0.124$\pm$  0.005      &  7.547 $\pm$0.291      &  8.105 $\pm$0.196      &  6.581  $\pm$0.292      &  6.566 $\pm$0.464     \\
   25F1      &    6.75$\pm$ 0.178  & 0.082$\pm$  0.002      &  7.624 $\pm$0.344      &  8.159 $\pm$0.223      &  6.828  $\pm$0.316      &   $\dots$     \\
    1F2      &   6.921$\pm$   0.127  & 0.082$\pm$  0.001      &  7.397  $\pm$0.260     &  8.398 $\pm$0.200      &  6.558  $\pm$0.284     &  5.997  $\pm$0.464     \\
    2F2      &  4.787$\pm$ 0.003 & 0.102$\pm$  0.006      &  7.678 $\pm$0.895     &   8.240 $\pm$0.579     &  6.846  $\pm$0.821     &  6.379 $\pm$0.795    \\
    8F2      &    5.676$\pm$    0.064  &  0.092$\pm$  0.002      &  7.574  $\pm$0/260      &  8.555 $\pm$0.176      &  6.642  $\pm$0.250      &  5.984  $\pm$0.452     \\
   11F2      &  7.082$\pm$  0.237 & 0.059$\pm$ 0.001      &  7.439  $\pm$0.319      &  8.776 $\pm$0.216     &  6.512  $\pm$0.307      &  6.195  $\pm$0.473     \\
   19F2      &    6.841$\pm$   0.339 &   0.070$\pm$ 0.001     &  7.745  $\pm$0.300      &  8.712$\pm$0.205      &  6.896  $\pm$0.294      &  6.325  $\pm$0.465     \\
   24F2    &   7.023$\pm$ 0.322& 0.070$\pm$ 0.001& \dots& 8.159$\pm$ 0.145& \dots& 6.702$\pm$0.500\\
     28F2      &   7.943$\pm$   0.418&    0.156$\pm$0.001 &            7.508  $\pm$0.303      &  7.877 $\pm$0.206     & 6.725$\pm$0.294&      6.188$\pm$0.467         \\
&&&&&&\\
\hline

\end{tabular}
\end{table*}

\begin{table*}

\caption{Radial metallicity gradients, and means}
\begin{tabular}{ccccccc}
\hline\hline
 {element}& {probe}& {sample}& {N}&  {slope}&  {intercept}&  {mean}\\
& & &&[dex kpc$^{-1}]$& log(X/H)+12& log(X/H)+12\\

\hline
&&&&&&\\

N& HII& GMOS,MMT& 27&  -0.067\tablefootmark{a} & 8.07& 7.62$\pm$0.048  \\
N& PNe& MMT\tablefootmark{b}& 14&   -0.049\tablefootmark{a}& 7.843&   7.545$\pm$0.059 \\

O& HII& GMOS,MMT& 28& -0.088$\pm$0.013\tablefootmark{c}& 9.202$\pm$0.107&  8.522$\pm$0.049 \\
O& PNe& MMT&17& -0.044$\pm$0.007\tablefootmark{c}& 8.59$\pm$0.062& 8.379$\pm$0.068\\

Ne\tablefootmark{d}& HII& MMT& 6& \dots& \dots& 7.979$\pm$0.117\\
Ne\tablefootmark{d}& PNe& MMT& 6& \dots &\dots& 7.813$\pm$0.103\\

Ar& HII& GMOS,MMT& 23& -0.037$\pm$0.016& 6.27$\pm$0.143& 6.21$\pm$0.057\\
Ar\tablefootmark{d}& MMT& PNe& 4& \dots & \dots& 6.077$\pm$0.144\\
&&&&&&\\
\hline
\end{tabular}
\tablefoot{
\tablefoottext{a}{{\it Fitexy} gives q=0, slope calculated with least square method instead.}
\tablefoottext{b}{Note that MMT samples of both PNe and HII regions are slightly different from those of Stanghellini et al. (2010), since two PNe have been newly classified as HII regions;}
\tablefoottext{c}{{\it Fitexy} gives q=0.051 for HII regions, 0.053 for PNe;}
\tablefoottext{d}{Sample on restricted domain, inadequate to calculate gradients.}
}

\end{table*}


\clearpage

\begin{thebibliography}{}
\bibitem[Baldwin et al.(1981)]{1981PASP...93....5B} Baldwin, J.~A., 
Phillips, M.~M., \& Terlevich, R.\ 1981, \pasp, 93, 5 
\bibitem[Bland-Hawthorn et al.(2010)]{2010ApJ...713..166B} Bland-Hawthorn, 
J., Krumholz, M.~R., \& Freeman, K.\ 2010, \apj, 713, 166 
\bibitem[Boissier 
\& Prantzos(1999)]{1999Ap&SS.265..409B} Boissier, S., \& Prantzos, N.\ 1999, \apss, 265, 409 
\bibitem[Bresolin et al.(2009)]{2009ApJ...695..580B} Bresolin, F., 
Ryan-Weber, E., Kennicutt, R.~C., \& Goddard, Q.\ 2009, \apj, 695, 580 
\bibitem[Bresolin et al.(2012)]{2012ApJ...750..122B} Bresolin, F., 
Kennicutt, R.~C., \& Ryan-Weber, E.\ 2012, \apj, 750, 122 
\bibitem[Cahn et 
al.(1992)]{1992A&AS...94..399C} Cahn, J.~H., Kaler, J.~B., \& Stanghellini, L.\ 1992, \aaps, 94, 399 
\bibitem[Chiappini et al.(1997)]{1997ApJ...477..765C} Chiappini, C., 
Matteucci, F., \& Gratton, R.\ 1997, \apj, 477, 765 
\bibitem[Cresci et al.(2010)]{2010Natur.467..811C} Cresci, G., Mannucci, 
F., Maiolino, R., et al.\ 2010, \nat, 467, 811 
\bibitem[Dav{\'e} et al.(2011)]{2011MNRAS.416.1354D} Dav{\'e}, R., 
Finlator, K., \& Oppenheimer, B.~D.\ 2011, \mnras, 416, 1354 
\bibitem[Davidge(2006)]{2006PASP..118.1626D} Davidge, T.~J.\ 2006, \pasp,
118, 1626
\bibitem[Ferland et al.(2013)]{2013RMxAA..49..137F} Ferland, G.~J., Porter, 
R.~L., van Hoof, P.~A.~M., et al.\ 2013, \rmxaa, 49, 137 
\bibitem[Finlator 
\& Dav{\'e}(2008)]{2008MNRAS.385.2181F} Finlator, K., \& Dav{\'e}, R.\ 2008, \mnras, 385, 2181 
\bibitem[Freedman et al.(2001)]{2001ApJ...553...47F} Freedman, W.~L., 
Madore, B.~F., Gibson, B.~K., et al.\ 2001, \apj, 553, 47 
\bibitem[Freeman 
\& Bland-Hawthorn(2002)]{2002ARA&A..40..487F} Freeman, K., \& Bland-Hawthorn, J.\ 2002, \araa, 40, 487 
\bibitem[Friel et al.(2002)]{2002AJ....124.2693F} Friel, E.~D., Janes, 
K.~A., Tavarez, M., et al.\ 2002, \aj, 124, 2693 
\bibitem[Garnett 
\& Shields(1987)]{1987ApJ...317...82G} Garnett, D.~R., \& Shields, G.~A.\ 1987, \apj, 317, 82 
\bibitem[Gibson et 
al.(2013)]{2013A&A...554A..47G} Gibson, B.~K., Pilkington, K., Brook, C.~B., Stinson, G.~S., \& Bailin, J.\ 2013, \aap, 554, A47 
\bibitem[Gottesman 
\& Weliachew(1975)]{1975ApJ...195...23G} Gottesman, S.~T., \& Weliachew, L.\ 1975, \apj, 195, 23 
\bibitem[Greenawalt et al.(1998)]{1998ApJ...506..135G} Greenawalt, B., 
Walterbos, R.~A.~M., Thilker, D., \& Hoopes, C.~G.\ 1998, \apj, 506, 135 
\bibitem[Hou et 
al.(2000)]{2000A&A...362..921H} Hou, J.~L., Prantzos, N., \& Boissier, S.\ 2000, \aap, 362, 921 
\bibitem[Immler \& Wang(2001)]{2001ApJ...554..202I} Immler, S., \& Wang, Q.~D.\ 2001, \apj, 554, 202 
\bibitem[Janes(1979)]{1979ApJS...39..135J} Janes, K.~A.\ 1979, \apjs, 39, 
135 
\bibitem[Jones et al.(2010)]{2010ApJ...725L.176J} Jones, T., Ellis, R., 
Jullo, E., \& Richard, J.\ 2010, \apjl, 725, L176 
\bibitem[Jones et al.(2013)]{2013ApJ...765...48J} Jones, T., Ellis, R.~S., 
Richard, J., \& Jullo, E.\ 2013, \apj, 765, 48 
\bibitem[Kingsburgh 
\& Barlow(1994)]{1994MNRAS.271..257K} Kingsburgh, R.~L., \& Barlow, M.~J.\ 1994, \mnras, 271, 257 
\bibitem[Kniazev et al.(2008)]{2008MNRAS.384.1045K} Kniazev, A.~Y., 
Pustilnik, S.~A., \& Zucker, D.~B.\ 2008, \mnras, 384, 1045 
\bibitem[Kobayashi 
\& Nakasato(2011)]{2011ApJ...729...16K} Kobayashi, C., \& Nakasato, N.\ 2011, \apj, 729, 16 
\bibitem[Kubryk et al.(2013)]{2013MNRAS.436.1479K} Kubryk, M., Prantzos, 
N., \& Athanassoula, E.\ 2013, \mnras, 436, 1479 
\bibitem[Kudritzki et al.(2012)]{2012ApJ...747...15K} Kudritzki, R.-P., 
Urbaneja, M.~A., Gazak, Z., et al.\ 2012, \apj, 747, 15 
\bibitem[Ma et al.(2005)]{2005PASP..117..256M} Ma, J., Zhou, X., Chen, J.,
et al.\ 2005, \pasp, 117, 256
\bibitem[Maciel 
\& Costa(2013)]{2013RMxAA..49..333M} Maciel, W.~J., \& Costa, R.~D.~D.\ 2013, \rmxaa, 49, 333 
\bibitem[Makarova et al.(2010)]{2010MNRAS.406.1152M} Makarova, L., Koleva, 
M., Makarov, D., \& Prugniel, P.\ 2010, \mnras, 406, 1152 
\bibitem[Magrini et 
al.(2001)]{2001A&A...379...90M} Magrini, L., Perinotto, M., Corradi, R.~L.~M., \& Mampaso, A.\ 2001, \aap, 379, 90 
\bibitem[Magrini et 
al.(2007)]{2007A&A...470..843M} Magrini, L., Corbelli, E., \& Galli, D.\ 2007, \aap, 470, 843 
\bibitem[Magrini et 
al.(2009)]{2009A&A...494...95M} Magrini, L., Sestito, P., Randich, S., \& Galli, D.\ 2009, \aap, 494, 95 (2009a)
\bibitem[Magrini et al.(2009)]{2009ApJ...696..729M} Magrini, L., 
Stanghellini, L., \& Villaver, E.\ 2009, \apj, 696, 729 (2009b)
\bibitem[Magrini et 
al.(2010)]{2010A&A...512A..63M} Magrini, L., Stanghellini, L., Corbelli, E., Galli, D., \& Villaver, E.\ 2010, \aap, 512, A63 
\bibitem[Maraston(2005)]{2005MNRAS.362..799M} Maraston, C.\ 2005, \mnras, 
362, 799 
\bibitem[Minchev et 
al.(2013)]{2013A&A...558A...9M} Minchev, I., Chiappini, C., \& Martig, M.\ 2013, \aap, 558, A9 
\bibitem[Moll{\'a} et al.(1997)]{1997ApJ...475..519M} Moll{\'a}, M., Ferrini, F., 
\& Diaz, A.~I.\ 1997, \apj, 475, 519 
\bibitem[Moll{\'a}(2014)]{2014arXiv1401.0761M} Moll{\'a}, M.\ 2014, 
arXiv:1401.0761 
\bibitem[Nantais et al.(2011)]{2011AJ....142..183N} Nantais, J.~B., Huchra,
J.~P., Zezas, A., Gazeas, K., \& Strader, J.\ 2011, \aj, 142, 183
\bibitem[Minchev et 
al.(2013)]{2013A&A...558A...9M} Minchev, I., Chiappini, C., \& Martig, M.\ 2013, \aap, 558, A9 
\bibitem[Osterbrock 
\& Ferland(2006)]{2006agna.book.....O} Osterbrock, D.~E., \& Ferland, G.~J.\ 2006, Astrophysics of gaseous nebulae and active galactic nuclei, 2nd.~ed.~by D.E.~Osterbrock and G.J.~Ferland.~Sausalito, CA: University Science Books, 2006
\bibitem[Patterson et al.(2012)]{2012MNRAS.422..401P} Patterson, M.~T., 
Walterbos, R.~A.~M., Kennicutt, R.~C., Chiappini, C., 
\& Thilker, D.~A.\ 2012, \mnras, 422, 401 
\bibitem[Park et al.(2012)]{2012ApJS..203....6P} Park, D., Kelly, B.~C., 
Woo, J.-H., \& Treu, T.\ 2012, \apjs, 203, 6 
\bibitem[Peimbert 
\& Torres-Peimbert(1983)]{1983IAUS..103..233P} Peimbert, M., \& Torres-Peimbert, S.\ 1983, Planetary Nebulae, 103, 233 
\bibitem[Perelmuter 
\& Racine(1995)]{1995AJ....109.1055P} Perelmuter, J.-M., \& Racine, R.\ 1995, \aj, 109, 1055 
\bibitem[Petit et 
al.(1988)]{1988A&AS...74..475P} Petit, H., Sivan, J.-P., \& Karachentsev, I.~D.\ 1988, \aaps, 74, 475 
\bibitem[Pilkington et 
al.(2012)]{2012A&A...540A..56P} Pilkington, K., Few, C.~G., Gibson, B.~K., et al.\ 2012, \aap, 540, A56 
\bibitem[Portinari 
\& Chiosi(1999)]{1999A&A...350..827P} Portinari, L., \& Chiosi, C.\ 1999, \aap, 350, 827 
\bibitem[Queyrel et 
al.(2012)]{2012A&A...539A..93Q} Queyrel, J., Contini, T., Kissler-Patig, M., et al.\ 2012, \aap, 539, A93 
\bibitem[Rahimi et al.(2011)]{2011MNRAS.415.1469R} Rahimi, A., Kawata, D., 
Allende Prieto, C., et al.\ 2011, \mnras, 415, 1469 
\bibitem[Ro{\v s}kar et al.(2008)]{2008ApJ...684L..79R} Ro{\v s}kar, R., 
Debattista, V.~P., Quinn, T.~R., Stinson, G.~S., 
\& Wadsley, J.\ 2008, \apjl, 684, L79 
\bibitem[Rupke et al.(2010)]{2010ApJ...723.1255R} Rupke, D.~S.~N., Kewley, 
L.~J., \& Chien, L.-H.\ 2010, \apj, 723, 1255 
\bibitem[Sanchez et al.(2013)]{2013arXiv1311.7052S} Sanchez, S.~F., 
Rosales-Ortega, F.~F., Iglesias-Paramo, J., et al.\ 2013, arXiv:1311.7052 
\bibitem[Sanders et al.(2012)]{2012ApJ...758..133S} Sanders, N.~E., 
Caldwell, N., McDowell, J., \& Harding, P.\ 2012, \apj, 758, 133 
\bibitem[Scarano \& L{\'e}pine(2013)]{2013MNRAS.428..625S} Scarano, S., \& L{\'e}pine, J.~R.~D.\ 2013, \mnras, 428, 625 
\bibitem[Sell et al.(2011)]{2011ApJ...735...26S} Sell, P.~H., Pooley, D.,
Zezas, A., et al.\ 2011, \apj, 735, 26

\bibitem[Sestito et 
al.(2008)]{2008A&A...488..943S} Sestito, P., Bragaglia, A., Randich, S., et al.\ 2008, \aap, 488, 943 
\bibitem[Stanghellini et al.(2003)]{2003ApJ...596..997S} Stanghellini, L., 
Shaw, R.~A., Balick, B., et al.\ 2003, \apj, 596, 997 
\bibitem[Stanghellini et al.(2008)]{2008ApJ...689..194S} Stanghellini, L., 
Shaw, R.~A., \& Villaver, E.\ 2008, \apj, 689, 194 
\bibitem[Stanghellini \& Haywood(2010)]{2010ApJ...714.1096S} Stanghellini, L., \& Haywood, M.\ 2010, \apj, 714, 1096 
\bibitem[Stanghellini et 
al.(2010)]{2010A&A...521A...3S} Stanghellini, L., Magrini, L., Villaver, E., \& Galli, D.\ 2010, \aap, 521, A3 
\bibitem[Stasi{\'n}ska et 
al.(2013)]{2013A&A...552A..12S} Stasi{\'n}ska, G., Pe{\~n}a, M., Bresolin, F., \& Tsamis, Y.~G.\ 2013, \aap, 552, A12 
\bibitem[Tumlinson et al.(2011)]{2011Sci...334..948T} Tumlinson, J., Thom, 
C., Werk, J.~K., et al.\ 2011, Science, 334, 948 
\bibitem[van Dokkum et al.(2010)]{2010ApJ...709.1018V} van Dokkum, P.~G., 
Whitaker, K.~E., Brammer, G., et al.\ 2010, \apj, 709, 1018 
\bibitem[Vila-Costas 
\& Edmunds(1992)]{1992MNRAS.259..121V} Vila-Costas, M.~B., \& Edmunds, M.~G.\ 1992, \mnras, 259, 121 
\bibitem[Werk et al.(2011)]{2011ApJ...735...71W} Werk, J.~K., Putman, 
M.~E., Meurer, G.~R., \& Santiago-Figueroa, N.\ 2011, \apj, 735, 71 
\bibitem[Wright(2006)]{2006PASP..118.1711W} Wright, E.~L.\ 2006, \pasp, 
118, 1711 
\bibitem[Yong et al.(2012)]{2012AJ....144...95Y} Yong, D., Carney, B.~W., 
\& Friel, E.~D.\ 2012, \aj, 144, 95 
\bibitem[Yuan et al.(2011)]{2011ApJ...732L..14Y} Yuan, T.-T., Kewley, 
L.~J., Swinbank, A.~M., Richard, J., 
\& Livermore, R.~C.\ 2011, \apjl, 732, L14 
\bibitem[Yun et al.(1994)]{1994Natur.372..530Y} Yun, M.~S., Ho, P.~T.~P., 
\& Lo, K.~Y.\ 1994, \nat, 372, 530 
\bibitem[Zaritsky et al.(1994)]{1994ApJ...420...87Z} Zaritsky, D., 
Kennicutt, R.~C., Jr., \& Huchra, J.~P.\ 1994, \apj, 420, 87 

\end{thebibliography}
\end{document}